\documentclass[11pt,a4paper]{article}
\usepackage{amstex,amsthm,amssymb,epsfig}

\def\tr{\operatorname{tr}}

\def\det{\operatorname{det}}

\newcommand{\bra}[1]{\langle\,#1\,|}
\newcommand{\ket}[1]{|\,#1\,\rangle}

\def\beqa{\begin{eqnarray}}
\def\eeqa{\end{eqnarray}}
\def\ba{\begin{array}}
\def\ea{\end{array}}
\def\r{\rangle}

\def\a{\alpha}

\def\la{\lambda}
\def\s{\sigma}
\def\sul{\sum\limits}
\def\pl{\prod\limits}
\def\lt({\left(}
\def\rt){\right)}
\def\pd #1{\frac{\partial}{\partial #1}}

\begin{document}

\begin{titlepage}
\begin{flushright}
\vphantom{Preprint number}%LPENSL-TH-15/99\\
\end{flushright}
\par \vskip .1in \noindent

\begin{center}
{\LARGE Correlation functions of the higher spin XXX chains}\\
\end{center}
  \par \vskip .3in \noindent

\begin{center}

      {\bf N. KITANINE$^{*}$}
  \par \vskip .1in \noindent

{\sl  Department of Mathematics\\
University of York\\
Heslington, York\\
YO10 5DD\\
UK
}\\[0.6in]
\end{center}

\par \vskip .10in
\begin{center}
{\bf Abstract}\\
\end{center}

\begin{quote}
Using the Algebraic Bethe Ansatz we consider the correlation functions of the integrable
higher spin chains. We apply a method recently developed for the spin $\frac 12$ Heisenberg
chain, based on the solution of the quantum inverse problem. We construct a representation
for the correlation functions on a finite chain for arbitrary spin. Then we show how the
string solutions of the Bethe equations can be considered in the framework of this approach
in the thermodynamic limit. Finally,  a multiple integral representation for the spin 1
zero temperature 
correlation functions is obtained in the thermodynamic limit.
 
\end{quote}

\vskip 1.4in

\begin{flushleft}
\rule{5.1 in}{.007 in}\\
$^{*}$ {\small On leave of absence from  the Steklov Mathematical Institute at 
St. Petersburg, Fontanka 27,\\ St Petersburg 191011, Russia.}\\

March 2001
\end{flushleft}

\end{titlepage}

\section*{Introduction}

A new method
of computation of the correlation functions and of the form factors of 
the Heisenberg spin $\frac 12$ chains
developed in \cite{KitMT99,IzeKMT99,kmt2} based on the
Algebraic Bethe Ansatz \cite{FadST79} and the resolution of the quantum
inverse problem \cite{KitMT99,mt} has given a possibility
to calculate a very large class of correlation functions for a finite chain and
in the thermodynamic limit. 
The zero temperature correlation functions (which are defined 
as mean values of some products of local operators with respect to the ground state 
of the model)  
were obtained \cite{kmt2} as multiple integrals which coincide for the case without
magnetic field
 with the results obtained using the vertex operators approach 
\cite{JimMMN92,JimM96}. However the
new method gave a better understanding of the structure of these results. 
It was shown,
in particular, 
 that the expressions under integrals 
can be
  separated  into two parts
with different origin: algebraic part which depends on  the choice  of local 
operators
and does not depend on the choice of the ground state
and analytic (or determinant) part which, on the contrary,
 is fixed uniquely by the ground state.
 
This very particular structure permits to hope that these results can be 
generalised 
for several more general situations in particular for the temperature 
dependent correlation 
functions for the Heisenberg spin $\frac 12$ 
but also for other integrable models with the same $R$-matrix, and first of
all for the higher spin chains (they can be considered as the first step
to the future generalisations).  These two apparently
very different problems have however one common detail: the main difficulty 
is the analysis 
of  excited states in the first case or of 
 more complicated ground states for the second one. In both cases we should deal 
with 
{\it bound states} or, more precisely, with the string solutions of the Bethe 
equations.
The understanding of the influence  of these bound states (quantum breathers) 
is a very important step to the calculation of the temperature and time 
dependent correlation
functions. 

For this reason, before considering a more complicated example of temperature 
dependent 
correlation functions, we consider the higher spin Heisenberg models. This case is 
more simple as the ground state of a higher spin XXX chain contains only strings of
one kind and not a mixture of different types of strings as an arbitrary excited 
state.
This problem is important also from other point of view as it can give some 
information
about other integrable models including integrable quantum field theories. 

%%%%%%%%%%%%%%%%%%%%%%%%%%%%%%%%%%%%%%%%%%%%%%%%%%%%%%%%%%%%%%%%%%%%%%%%%%%
%%%%%%%%%%%%%
In this paper we consider the correlation functions of the XXX higher spin 
chains. 
This model was first considered by Kulish, Sklyanin and Reshetikhin 
in \cite{KulRS} where the notion of {\it fusion} was introduced. 
It  was solved by means of the Algebraic Bethe Ansatz by L. Takhtajan 
\cite{takh} and independently by H. Babujian in \cite{bab},
 the thermodynamics of this model was considered in \cite{bab}. 
The XXZ version of higher spin chains was introduced in \cite{FatZ} 
and solved in \cite{KirR},
but these models are not considered here for several reasons. The correlation 
functions
of the XXZ spin 1 model in the anti-ferromagnetic regime were calculated  
in \cite{Idz,BogW,Kon}
using the vertex operator approach. 
We propose here a different way of calculation of the 
correlation functions based on the Algebraic Bethe Ansatz.

 As in \cite{kmt2} the first
step of computation of the correlation functions is the  solution of the 
{\it quantum inverse
problem}. Such a solution for a very large class of quantum integrable models
 including the higher spin Heisenberg chains was recently obtained \cite{mt}
in a form very similar to the spin $\frac 12$ case. Taken together with the results 
for the scalar products of the Bethe states \cite{Gau67,Kor82,Sla89} it permits to obtain a
 representation
for the finite chain correlation functions for arbitrary spin. On this stage one 
should
take the thermodynamic limit and, hence,
introduce the string solutions. We illustrate this procedure using the simplest
 example of
the spin 1 chain. We would like to underline that a similar procedure is possible
 also for 
higher spin models but it leads to  more cumbersome calculations and we present only one
  form
of the  result for this case without detailed derivation.

The main difficulty which arises from the presence 
of bound states is the fact that the
 algebraic part
becomes singular.  This problem can be
 solved by 
chosing  the integration contours in the multiple integral representations taking
into account the sign of the finite size corrections to the string picture \cite{dVW,kb,kbp}. 
After this modification
one can see that two parts appear again and the determinant part once again 
is defined
uniquely by the ground state. Such a result is given in section 5 of this paper. %for
%the spin 1 chain.% Similar results can be also obtained for spin $\frac 32$. 
%However up to now
%there is no general proof for such representations for arbitrary spin. 

 The main result 
of this paper is the fact that the mean values with respect to the states
containing bound states can be calculated in the framework of our approach.
 It means
in particular that some new tools introduced here can be used also to
calculate the temperature dependent correlation functions.

This paper is organised as follows. In the first section we
 introduce the higher spin Heisenberg
chains following the papers  of Takhtajan \cite{takh} 
and Babujian \cite{bab}. 
The solution of the inverse problem for 
these models \cite{mt} is given in the second section. 
This solution is used to obtain 
representations for the correlation functions on a
 finite chain for arbitrary spin in 
section 3, we show, in particular, how to override 
the additional algebraic difficulties
appearing in the higher spin case. The thermodynamic 
limit of the spin 1 chain is considered
in two last sections. We show how to deal with 2-strings 
in the thermodynamic limit using the 
simplest example of one point functions in section 4. 
This first example permits two elaborate
some simple rules which permit to deal with strings for general correlation functions
(section 5).

\section{XXX Heisenberg chain with arbitrary spin}
\setcounter{equation}{0}

In this section we introduce the XXX Heisenberg chains for arbitrary spin.
 We follow
in general here the papers of Takhtajan \cite{takh} and
 Babujian \cite{bab}. 

Unlike the spin $\frac 12$ case we start directly 
from the $L$-operator and construct
later the Hamiltonian from the transfer matrix. It is necessary 
to obtain an integrable
generalisation of the usual Heisenberg chain (a direct generalisation of the 
spin  
$\frac 12$ XXX Hamiltonian is not integrable for higher spins).
 However the $L$-operator 
can be obtained by the direct generalisation:

\begin{equation}
L_m(\la)= \frac{1}{\la-i(s+\frac 12)}\left(\ba{cc}\la- 
i (s_m^z+ \frac 12) & -i s_m^-\\
                             -is^+_m  &\la+ i( s_m^z+\frac 12)\ea\right)
\end{equation}
One should note that for this $L$-operator the auxiliary
 space is two-dimensional but the quantum space
has $2s+1$ dimensions. The matrices $s^z$, $s^\pm$ are the
 spin operators in the representation
 of spin $s$.
This  $L$- operator has the same intertwining relation with
 the rational $4\times 4$ $R$ matrix
\begin{equation}
   R(\la)=
  \left(\ba{cccc}1&0&0&0\\
                 0&\frac{\la}{\la-i}&\frac {-i}{\la-i}&0\\
                 0&\frac {-i}{\la-i}&\frac{\la}{\la-i}&0\\
                 0&0&0&1\ea\right),
\end{equation}
as in the spin $\frac 12$ case:

\begin{equation} \label{rll}
   R_{12}(\lambda- \mu)\ (L_m)_1(\lambda)\ (L_m)_2(\mu)  
      = (L_m)_2(\mu)\ (L_m)_1(\lambda)\ R_{12}(\lambda- \mu).
\end{equation}
As usual, indices $1$ and $2$ in $L$ operators denote two different 
auxiliary spaces. 

The next step is the construction of the monodromy matrix for a spin chain 
of $M$ sites:
\begin{equation*}
T(\la)=L_M(\la-\xi_M)L_{M-1}(\la-\xi_{M-1})
\dots L_1(\la-\xi_1)\equiv\left(\ba{cc}A(\la)&B(\la)\\
                                       C(\la)&D(\la)\ea\right),
\end{equation*}
with arbitrary inhomogeneity parameters $\xi_j$.  For this monodromy
 matrix one can also write 
the ``commutation relation'' with $R$-matrix:
\begin{equation} \label{rtt}
   R_{12}(\lambda- \mu)\ T_1(\lambda)\ T_2(\mu)  
      = T_2(\mu)\ T_1(\lambda)\ R_{12}(\lambda- \mu),
\end{equation}
 and it contains the commutation relation for the operators
 $A(\la)$, $B(\la)$, $C(\la)$ and $D(\la)$ 
acting in the quantum space.

Up to this point all the construction was 
the same as in the spin $\frac 12$ case but to construct
 local Hamiltonians from the monodromy matrix one should 
introduce some new concepts. The trace
identities for the  spin $\frac 12$ case were based on 
the fact that $L_n(0)$ is just a transposition
operator in the tensor product of the auxiliary and quantum spaces. 
Now the quantum and auxiliary spaces
have different numbers of dimensions. Hence it is necessary to 
construct a monodromy matrix with a
$2s+1$ dimensional auxiliary space. Such a construction  
was proposed by Kulish, Sklyanin and 
Reshetikhin \cite{KulRS} and it is called {\it  fusion}.

The fused $L$-operator $L^{(s)}_n(\la)$ can be constructed 
as a following projection on the symmetric subspace
in the tensor product of  $2s$ auxiliary spaces (this symmetric
 subspace has exactly $2s+1$
dimensions) of the following product of local $L$-operators:
\begin{equation}
L_n^{(s)}(\la)=P^+_{a_1,\dots a_n}L_{a_{2s},n}(\la+2is-i)\dots 
L_{a_2,n}(\la+i) L_{a_1,n}(\la)
P^+_{a_1,\dots a_n}
\end{equation} 
where indices $a_j$ mark the corresponding auxiliary spaces 
and $P^+_{a_1,\dots a_n}$ is
the symmetrizer (projector on the symmetric subspace).
 Thus we constructed the $L$-operator
with $2s+1$-dimensional auxiliary space and satisfying the following property:
\[L^{(s)}_{a,n}(-is)=P_{a,n},\]
where $P_{a,n}$ is the transposition in the tensor product of 
the auxiliary space and 
local quantum space.This property is crucial for the construction of
   local Hamiltonians 
\cite{bab}
and for the solution of the inverse problem \cite{mt}. 
The fused monodromy matrix is 
constructed as usual:
\[T^{(s)}_a(\la)=L^{(s)}_M(\la-\xi_M)L^{(s)}_{M-1}(\la-\xi_{M-1})
\dots L^{(s)}_1(\la-\xi_1)\]
The fused transfer matrix which is the trace of the monodromy
 matrix taken in the auxiliary space,
\[\tau^{(s)}(\la)=\tr_a T_{a}(\la)\]
 commutes not only with the transfer matrices for any value of 
parameter $\la$ but also
with the ``ordinary'' transfer matrix $A(\la)+D(\la)$ (as it is a 
polynomial function of  
$A(\la)+D(\la)$):
\begin{equation}
[\tau^{(s)}(\la),\tau^{(s)}(\mu)]=[\tau^{(s)}(\la),A(\mu)+D(\mu)]=0.
\end{equation}
It means, in particular, that the Hamiltonians constructed from the
 fused transfer matrix 
in the homogeneous case $\xi_j=0$ 
using the trace identities:
\begin{equation}
H^{(s)}=\mathrm{const}\left.\frac d{d\la}\tau^{(s)}(\la)\right|_{\la=-is},
\label{trace}
\end{equation}
also commute with the ``ordinary'' transfer matrix and can be diagonalized by the 
{\it Algebraic Bethe Ansatz} \cite{FadST79} procedure.
The Hamiltonians constructed by trace identities are local, translation invariant
 and can be written as polynomials of degree $2s$ of the local spin-spin
 interaction terms 
\cite{takh}:
\begin{align}
H^{(s)}=&\sul_{m=1}^M Q_{2s}(\mathbf{s}_m\mathbf{s}_{m+1}),\\
Q_{2s}(x)=&\sul_{j=1}^{2s}\left(\sul_{k=1}^j\frac 1k\right)
\pl_{l=0\atop{l\neq j}}^{2s}
\frac{x-x_l}{x_j-x_l},
 \end{align}
where $\mathbf{s}_n=(s_n^x,s_n^y,s_n^z)$ are spin operators and
 $x_l=\frac 12 [l(l+1)-2s(s+1)]$.
The first example of such a Hamiltonian is the spin 1 case where
\begin{equation}
H^{(1)}=\frac 14 \sul_{m=1}\mathbf{s}_m\mathbf{s}_{m+1}-
(\mathbf{s}_m\mathbf{s}_{m+1})^2.
\end{equation}

As we already mentioned to diagonalize this Hamiltonian one can 
use the usual Algebraic
Bethe Ansatz. We start from the ferromagnetic state $\ket{0}$ with all the spins up
\[s_n^+\ket{0}=0,\quad \forall n\]
which is an eigenstate of the Hamiltonians (\ref{trace})  and consider the action of
the generalised creation operators $B(\la)$ on this state. It is easy to see that
a state 
\[B(\la_1)B(\la_2)\dots B(\la_N)\ket{0}\]
 is an eigenstate of the ordinary transfer matrix
\begin{equation}
(A(\mu)+D(\mu))B(\la_1)B(\la_2)\dots B(\la_N)\ket{0}=\tau(\mu,\{\la_j\})
B(\la_1)B(\la_2)\dots B(\la_N)\ket{0},
\label{eigentau}
\end{equation}
and, hence, of the Hamiltonians if the  parameters $\{\la_j\}$ satisfy the following
{\it Bethe equations}
\begin{equation}
\varphi_j(\{\la\})\equiv
\left(\frac{\la+is}{\la-is}\right)^M
\pl_{k=1}^N\frac{\la_j-\la_k-i}{\la_j-\la_k+i}=-1.
\label{bethe}
\end{equation}

These Bethe states have exactly the same property as 
the Bethe states for the spin $\frac 12$
case in particular one can prove the Gaudin formula 
\cite{Gau67,Kor82} for their norms 
\begin{align}
\label{gaud}
\bra{0}\pl_{j=1}^NC(\la_j)\pl_{k=1}^N B(\la_k)\ket{0}=&(-1)^N\pl_{j\neq k}
\frac{\la_j-\la_k-i}{\la_j-\la_k}\det_N
\Phi'(\{\la\}),\\
\Phi'(\{\la\})_{a b}=&-i\pd{\la_b}\log\varphi_a(\{\la\}). \nonumber
\end{align}
and Slavnov formula \cite{Sla89,KitMT99} for  scalar products of a Bethe state 
$\pl_{k=1}^N B(\la_k)\ket{0}$ and a state $\bra{0}\pl_{j=1}^NC(\mu_j)$ with 
arbitrary set
of parameters: 
\begin{align}
\label{slav}
\bra{0}\pl_{j=1}^NC(\mu_j)\pl_{k=1}^N B(\la_k)\ket{0}=&\frac{\det_N 
T(\{\la,\mu\}) }
{\det_N V(\{\la,\mu\})},\\
T(\{\la,\mu\})_{a b}=&\pd{\la_a}\tau(\mu_b,\{\la\}), \qquad 
V(\{\la,\mu\})_{a b}=\frac 1{\la_a-\mu_b},\nonumber
\end{align}
where $\tau(\mu_b,\{\la\})$ is the eigenvalue of the ordinary 
transfer matrix (\ref{eigentau}).

The thermodynamic limit for the higher spin Heisenberg model is slightly
 more complicated
than in spin $\frac 12$ case.
The ground state of the spin $s$ XXX model in the 
thermodynamic limit can be constructed using the ``string'' solutions
of the Bethe equations. It was shown in \cite{takh} that for the spin
$s$ chain the ground state contains only strings of length $2s$. 
Such a string solution can be written as follows
\[\la_k^a=\mu_k+i(a-s-\frac 12),\]
where $\mu_k$ is real and called the string centre. The particularity of the 
ground state is
the fact that it contains only  strings of one particular length.

The density of string centres in the ground state can be obtained from the 
following
 integral equation 
similar to the Lieb equation in spin $\frac 12$ case:
\begin{equation}
\rho_{2s}(\la)+2\sul_{a=1}^{2s-1} \int\limits_{-\infty}^{\infty}d\mu 
K_a(\la-\mu)\rho_{2s}(\mu)+\int\limits_{-\infty}^{\infty}d\mu 
K_{2s}(\la-\mu)\rho_{2s}(\mu)=\sul_{k=1}^{2s} K_{s}(\la+i(s+\frac 12-k)),
\end{equation}
where the kernels $K_j(\la)$ are defined as
\begin{equation}
K_a(\la)=\frac{2a}{(\la+ia)(\la-ia)}
\label{kernel}
\end{equation}
   The solution of this equation can be easily obtained
\begin{equation}
\rho_{2s}(\la)=\frac{1}{2\cosh (\pi \la)}.
\end{equation}
It is a remarkable property of the spin chains that for any spin the density is 
the same
as in the spin $\frac 12$ case, but one should note here that the ground states
 are however
very different being constructed from the different types of strings.

These results  for thermodynamics of 
the spin chains of arbitrary spin obtained by L. Takhtajan \cite{takh} and 
H. Babujan in  \cite{bab}
will be used in next sections for the calculation of the correlation functions.

\section{Inverse problem}
\setcounter{equation}{0}

In this section we remind the solution of the inverse scattering 
problem for the spin $s$ XXX chain. In other words
we reconstruct the local spin operators in terms of the fused monodromy
matrices. We follow the approach proposed by J.-M. Maillet and V. Terras
in \cite{mt}. 

 To illustrate the results of \cite{mt} we start with the spin 1 chain. In this case
 the fused monodromy matrix
has the following form:

\begin{equation*}
T^{(2)}(\la)=\left( \ba{ccc}
A(\la+i)A(\la) & \frac {1}{\sqrt{2}}(A(\la+i)B(\la)+B(\la+i)A(\la))& 
B(\la+i)B(\la)\\
&&\\
\frac {1}{\sqrt{2}}(A(\la+i)C(\la)+&
\frac 12 (A(\la+i)D(\la)+D(\la+i)A(\la)+&\frac {1}{\sqrt{2}}(D(\la+i)B(\la)+\\
+C(\la+i)A(\la))&+B(\la+i)C(\la)+
C(\la+i)B(\la))&+B(\la+i)D(\la))\\
&&\\
C(\la+i)C(\la)& \frac {1}{\sqrt{2}}
(D(\la+i)C(\la)+C(\la+i)D(\la))&A(\la+i)A(\la)
\end{array}\right)
%\label{fus1}
\end{equation*}

This matrix can be used to construct the operators $s^z_n$ and $s^\pm_n$, 
but here we are mostly
interested in the reconstruction of the local elementary matrices 
\[\left(E^{\a',\a}\right)_{a b}=\delta_{\a' a}\delta_{\a b}\] 
The local operators $E^{\a'_j,\a_j}_j$ can be expressed in terms of the 
corresponding monodromy matrix elements:

 \begin{equation}
E^{\a'_m,\a_m}_m=\pl_{j=1}^{m-1}\tau^{(2)}(\xi_j-i)\,
 T^{(2)}_{\a_m,\a'_m}(\xi_m-i)\,
\pl_{j=m+1}^{M}\tau^{(2)}(\xi_j-i),
\end{equation}
where $\tau^{(2)}(\la)=\tr_0\,T^{(2)}(\la)$ is the fused transfer matrix.

This result can be easily generalised for an arbitrary  spin $s$. 
 Now the fused monodromy matrix is a $(2s+1)\times(2s+1)$ matrix
and its elements are again  sums of different products of $2s$ fundamental 
monodromy matrix elements:

\begin{equation}   
T^{(2s)}_{\a,\a'}(\la)=\frac 1{(C_{2s}^{\a-1}C_{2s}^{\a'-1})^{\frac 12}}
\sul_{j_1+\dots j_{2s}-2s=\a-1\atop{k_1+\dots k_{2s}-2s=\a'-1}}
%\,\,\sul_{k_1+\dots k_{2s}-2s=\a'-1}
\!\!\!\!\!\!\!\!
T_{j_{2s}k_{2s}}(\la+2si-i)\dots T_{j_2k_2}(\la+i)T_{j_1k_1}(\la),
\label{Sinv}
\end{equation}
where $C_n^k$ are binomial coefficients.
For example the corner matrix element $T^{(2s)}_{1,1}(\la)$ is just a product 
of $2s$ operators $A(\la-ki)$, $k=0,\dots,2s-1$. 

To reconstruct the elementary local operators one should again dress these
 monodromy matrix elements by the corresponding fused transfer matrices:
 \begin{equation}
E^{\a'_m,\a_m}_m=\pl_{j=1}^{m-1}\tau^{(2s)}(\xi_j-is)\,
 T^{(2s)}_{\a_m,\a'_m}(\xi_m-is)\,
\pl_{j=m+1}^{M}\tau^{(2s)}(\xi_j-is),
\end{equation}

 The shifts of the inhomogeneity parameters are chosen in such a way that the 
eigenvalue of the fused
transfer matrix taken in the points $\xi_j+is$ on a Bethe state is:
 \begin{equation}
\tau^{(2s)}(\xi_j-is)B(\la_1)\,\dots B(\la_N)\ket{0}
=\pl_{k=1}^N\frac{\la_k-\xi_j-is}
                 {\la_k-\xi_j+is}\,\,B(\la_1)\dots B(\la_N)\ket{0}.
 \end{equation}

This is the solution of the quantum inverse problem for the arbitrary spin
 Heisenberg 
chain. Now we can use these representation to calculate the correlation 
functions and 
to do it we should first of all understand how these complicated combination of
the fundamental monodromy matrix elements act on Bethe states.

\section{Finite lattice correlation functions}
\setcounter{equation}{0}
\label{actions}

As our ultimate goal is to calculate the mean values of products of local 
operators with respect to  a Bethe state and, in particular to the ground state, 
following
the same ideas as in \cite{kmt2} we should
consider the action of the local operators on Bethe states. As the Bethe states
are eigenstates of the fused transfer matrix we need only to consider the action
of the elements of the fused monodromy matrix. The result of this action is the
``algebraic part'' of the final expression for the correlation functions.
We begin by considering   the action of a single local
operator on a Bethe state.

First of all we remind the action of the operators $A$, $B$ and $D$ (elements
of the fundamental monodromy matrix) on a
``bra'' Bethe state:

\begin{align}
 \bra{0}\pl_{k=1}^{N}C(\la_k)\,A(\la_{N+1})=&
  \sul_{a'=1}^{N+1} a(\la_{a'}) 
   \frac{\pl_{k=1}^N   
   (\la_k-\la_{a'}-i)}{\pl_{k=1\atop{k\neq a'}}^{N+1}(\la_k-\la_{a'})}
   \,\bra{0}\pl_{k=1\atop{k\neq a'}}^{N+1} C(\la_k);\label{abbb}\\
\bra{0}\pl_{k=1}^{N}C(\la_k)\, D(\la_{N+1})=&
  \sul_{a=1}^{N+1} d(\la_a) 
   \frac{\pl_{k=1}^N   
   (\la_a-\la_k-i)}{\pl_{k=1\atop{k\neq a}}^{N+1}(\la_a-\la_k)}
   \,\bra{0}\pl_{k=1\atop{k\neq a}}^{N+1} C(\la_k)
   ;\label{dbbb}\end{align}
%The action of the operator $B(\la)$ is more complicated:
\begin{align}
\bra{0}\pl_{k=1}^N C(\la_k)\, B(\la_{N+1})=&\sul_{a=1}^{N+1} 
  d(\la_a)\frac{\pl_{k=1}^N   
   (\la_a-\la_k-i)}{\pl_{k=1\atop{k\neq a}}^{N+1}(\la_a-\la_k)}
\times\nonumber\\
\times&\sul_{a'=1\atop{a'\neq a}}^{N+1} \frac{a(\la_{a'})}{(\la_{N+1}-\la_{a'}-i)}
\frac{\pl_{j=1\atop{j\neq a}}^{N+1}   
   (\la_j-\la_{a'}-i)}{\pl_{j=1\atop{j\neq a,a'}}^{N+1}(\la_j-\la_{a'})}
 \bra{0} \!\!\pl_{k=1\atop{k\neq a,a'}}^{N+1}\!\! C(\la_k). \label{cbbb}
\end{align}
It can be seen from these formulae that there are two 
different type of sums produced by
the action of the monodromy matrix elements 
``$A$-type'' and ``$D$-type'', (action of the operator 
$B$ produce for example one ``$A$-type'' sum and one  ``$D$-type'' sum). 
In our case the
eigenvalues of the operators $A(\la)$ and $D(\la)$ in the ferromagnetic
state are
\[a(\la)=1,\quad d(\la)=\pl_{j=1}^M\left(\frac{\la-\xi_j+is}{\la-\xi_j-is}\right)\]

One should note that in order to calculate the correlation functions we should
act by the ``strings'' of operators. Consider first the action of one local 
operator $E^{\a,\a}_m$. 
As it was shown 
in the previous section it can be written as a sum of ordered products
 of the fundamental monodromy 
matrix elements taken in the points $\xi-is,\xi-is+i,\dots, \xi+is-i$.
 The monodromy matrix
elements acting on the ground state produce the sums on the ``$D$-type'' 
indices $a_j$
and ``$A$-type'' indices $a'_j$ and the number of such sums is the same
for all the products corresponding to a fixed local operator, namely $\a-1$ 
``$D$-type'' sums and $2s-\a'+1$ ``$A$-type'' sums. Introducing the new notations
$\la_{N+j}=\xi-i(s-j+1)$  we can just repeat the 
calculations for the spin $\frac 12$ \cite{kmt2}
 taking however
into 
account that now $d(\la_{N+j})\neq 0$ for $j>1$:  
\begin{align*}
\frac{\bra{0}\pl_{k=1}^N C(\la_k)\,
\Big(\tau^{(2s)}(\xi-is)\Big)^{-1}\,\pl_{j=1}^{2s}T_{\a_j,\a'_j}
(\la_{N+j})
\,\pl_{k=1}^N B(\la_k)\ket{0}}
{\bra{0}\pl_{k=1}^N C(\la_k)\,\pl_{k=1}^N B(\la_k)\ket{0} }=&\\
=\frac{1}{\pl_{k=1}^{2s-1} i^k\, k!}
\sul_{a_j,a'_j=1}^{N+2s}H_{\{a_j,a'_j\}}(\la_1,\dots,\la_{N+2s})
\frac{\det \Psi(\{a,a'\})}{\det\Phi'},&
\end{align*}
where the function $H$ is defined as 
\begin{align}   
H_{\{a_j,a'_j\}}(\{\la\})=&
\frac{(-1)^{2s} }{\pl_{k>l}
(\la_{b_k}-\la_{b_l}-i)}\pl_{j=1}^{2s-\a+1}\left(\pl_{k=1}^{j-1}
(\la_{a'_j}-\la_{N+k}+i)\pl_{k=j+1}^{2s}(\la_{a'_j}-\la_{N+k})\right)
\times\nonumber\\
\times&
\pl_{j=1}^{\a-1}d(\la_{a_j})\pl_{k=1}^N
\frac{\la_{a_j}-\la_{k}-i}{\la_{a_j}-\la_{k}+i}
\left(\pl_{k=1}^{j-1}
(\la_{a_j}-\la_{N+k}-i)\pl_{k=j+1}^{2s}(\la_{a_j}-\la_{N+k})\right),
\label{underint}
\end{align}
For the indices $a_j$, $a'_j$, $b_j$ etc we conserve the notation 
of the spin $\frac 12$ case:
\[\{b_1,\dots,b_m\}=\{a'_{2s-\a+1},\dots,a'_{1},
a_{1},\dots,a_{\a-1}\}.\]
The determinant in the denominator is the Gaudin determinant and the matrix in the 
numerator is also
the Gaudin matrix with some replaced columns. We will consider this 
``analytic part'' in general
later on in
this section, and for the ground state for the spin 1 case in the last sections. 
One can easily obtain  representations for the replaced columns from the 
scalar product formula.
Here we will consider the ``algebraic part'' of the expression (\ref{underint}). 
Taking into account 
values of the parameters $\la_{N+k}$ we obtain:
\begin{align}   
H_{\{a_j,a'_j\}}(\{\la\})=&
\frac{(-1)^{\a-1} }{\pl_{k>l}
(\la_{b_k}-\la_{b_l}-i)}\pl_{j=1}^{\a-1}d(\la_{a_j})\pl_{k=1}^N\frac{\la_{a_j}-\la_{k}-i}
{\la_{a_j}-\la_{k}+i}
\left(\pl_{k=1}^{2s-1}
(\la_{a_j}-\xi+i(s-k))\right)\times\nonumber\\
\times&\pl_{j=1}^{2s-\a+1}\left(\pl_{k=1}^{j-1}
(\la_{a'_j}-\xi+i(s-k+2))\pl_{k=j+1}^{2s}(\la_{a'_j}-\xi+i(s-k+1))\right)
%\times\nonumber\\
%\times&\,\, 
\label{underint1}
\end{align}
One can see here that the product corresponding to the 
``$D$-type'' parameters is the same for any 
element of the sum in (\ref{Sinv}). Moreover from this 
result one can see that $a_j>N$ gives a non zero
contribution only if $a_j>N+1$ and there is $a_k=a_{j}-1$, $k<j$. 
It leads to the conclusion that
such a term should contain $d(\xi-is)$ which is zero. 
It means that as in spin $\frac 12$ case the 
summations over $a_j$ should be taken only from $1$ to $N$.

 The product corresponding to the ``$A$-type'' parameters is
not the same for all the terms 
but taking the sum in (\ref{Sinv}) and symmetrizing over the  
permutations of the ``$D$-type'' and 
``$A$-type'' parameters separately one can simplify it and obtain finally:
\begin{align}
f_{\a}(1,s)\equiv &\frac{\bra{0}\pl_{k=1}^N C(\la_k)\,E^{\a,\a}_m
\,\pl_{k=1}^N B(\la_k)\ket{0}}
{\bra{0}\pl_{k=1}^N C(\la_k)\,\pl_{k=1}^N B(\la_k)\ket{0} }=\nonumber\\
=&(-1)^{\a-1} \frac{C_{2s}^{\a-1}}{\pl_{k=1}^{2s-1} i^k\, k!}
\sul_{a_j=1}^N\sul_{a'_j=1}^{N+2s}\mathbf{H}^\a_{\{a_j,a'_j\}}(\la_1,\dots,\la_{N+2s})
\frac{\det \Psi(\{a,a'\})}{\det\Phi'},
\label{onepsumm}
\end{align}
\begin{align}
\mathbf{H}^\a_{\{a_j,a'_j\}}(\{\la\})=&
\pl_{ k,l=1\atop{k>l}}^{\a-1}
\frac{\la_{a_k}-\la_{a_l}}{(\la_{a_k}-\la_{a_l})^2+1}\pl_{ k,l=1\atop{k>l}}^{2s-\a+1}
\frac{\la_{a'_k}-\la_{a'_l}}{(\la_{a'_l}-\la_{a'_k})^2 +1}
\pl_{ k=1}^{\a-1}\pl_{l=1}^{2s-\a+1}\frac{1}
{\la_{a_k}-\la_{a'_l}-i}\times\nonumber\\ 
\times&\pl_{j=1}^{\a-1}
\left(\pl_{k=1}^{2s-1}
(\la_{a_j}-\xi+i(s-k))\right)
\pl_{j=1}^{2s-\a+1}\left(\pl_{k=1}^{2s-1}
(\la_{a'_j}-\xi+i(s-k+1))\right)
%\times\nonumber\\
%\times&\,\, 
\label{underint2}
\end{align}
One can see from this representation that terms with $a'_j>N$ produce  non zero 
contributions  if
$a'_j=N+2s$ or if there is $a'_k=a'_j +1$. It means that the operators
$C(\xi-i(s-k))$ which appear in the scalar 
product after the action of the local operators
should form a ``substring'' without holes starting from $\xi-is$.
 For example  states like
\[\bra{0}C(\xi-is)C(\xi-i(s-1))
\dots C(\xi-i(s-k))\pl_{b\le N\atop{b\neq a_j,a'_j}}C(\la_b)\]
produce non zero contributions to the correlation
 functions but the contribution of such
states as
  \[\bra{0}C(\xi-is)C(\xi-i(s-2))\!\!\!\!
\pl_{b\le N\atop{b\neq a_j,a'_j}}\!\!\!\!C(\la_b)
\,\,\,\,\,\,\text{or}\,\,\,\,\,\,
\bra{0}C(\xi-i(s-1))\!\!\!\!\pl_{b\le N\atop{b\neq a_j,a'_j}}\!\!\!\!C(\la_b)\]
are zero. This property is rather important as the matrix appearing
 in the scalar product is
much simpler in this case. 

Consider the determinant $\det\Psi$ appearing in (\ref{underint}) 
 from the scalar product 
\[\bra{0}C(\xi-is)C(\xi-i(s-1))\dots C(\xi-i(s-k+1))\!\!\!\pl_{b=k+1}^{N}\!\!\!
C(\la_b)\,\pl_{a=1}^N B(\la_a)\ket{0}\]
After extracting of the normalisation coefficients we obtain the following matrix:
\begin{alignat*}{2}
\Psi_{a b}=&\Phi'_{ab},\qquad &&b>k,\\
\Psi_{a 1}=&\frac 1{(\la_a-\xi+is)(\la_a-\xi+i(s-1))},&&b=1,\\
\Psi_{a b}=&\frac 1{(\la_a-\xi+i(s-b+1))(\la_a-\xi+i(s-b))}+ \\
&+
f_b\frac1{(\la_a-\xi+i(s-b+2))(\la_b-\xi+i(s-b+1))},\qquad &1<&b\le k.
\end{alignat*}
where
\[f_b=d(\xi-i(s-b+1))\pl_{j=1}^N\frac{\la_j-\xi+i(s-b+2)}{\la_j-\xi+i(s-b)}\]
The columns with $1<b\le k$ can be considered as  sum's of two columns and
 the second one does not 
contribute to   the determinant as it is always linearly dependent on the 
 columns with $a'<a$.
Thus the matrix $\Psi$ can be replaced by $\tilde{\Psi}$
\begin{alignat*}{2}
\tilde{\Psi}_{a b}=&\Phi'_{ab},\qquad &b>k&,\\
\tilde{\Psi}_{a b}=&\frac 1{(\la_a-\xi+i(s-b+1))
(\la_a-\xi+i(s-b))}\equiv p_b'(\la_a-\xi),\qquad &b\le k &.
\end{alignat*} 

A similar calculations can be done also for  $m$-point functions  
leading to the following
representation: 
\begin{align}
f_{\{\a,\a'\}}(m,s)\equiv &\frac{\bra{0}\pl_{k=1}^N C(\la_k)\,\pl_{l=1}^m 
E^{\a'_l,\a_l}_l
\,\pl_{k=1}^N B(\la_k)\ket{0}}
{\bra{0}\pl_{k=1}^N C(\la_k)\,\pl_{k=1}^N B(\la_k)\ket{0} }=\nonumber\\
= &(-1)^{\sum(\a'_l-1)}
\frac{\left(\pl_{l=1}^m C_{2s}^{\a_l-1}C_{2s}^{\a'_l-1}\right)^{\frac 12}}{
\left(\pl_{k=1}^{2s-1} i^k\, k!\right)^m \pl_{j,k=1\atop{j>k}}^m\pl_{r=1}^{2s}
\pl_{n=1}^{2r-1}
\Big(\xi_j-\xi_k-i(r-n)\Big)}\times\nonumber\\
&\times
\sul_{a_{jl}=1}^N\sul_{a'_{jl}=1}^{N+2sm}
\mathbf{H}^\a_{\{a_{jl},a'_{jl}\}}(\{\la\})
\frac{\det \Psi(\{a,a'\})}{\det\Phi'}.
\label{mpointsum}
\end{align}
 Here we introduced some new notations. In every site $l$
there is a local operator $E^{\a'_l,\a_l}_l$ which produces 
 sums over $\a_l-1$ ``$D$-type'' indices 
$a_{j l}$ and $2s+1-\a'_l$ ``$A$-type'' indices $a'_{j l}$, and we define
$\la_{N+2sl+k}=\xi_l-i(s-k+1)$.  We obtain the algebraic part: 
\begin{align}
\mathbf{H}^{\a,\a'}_{\{a_{j l},a'_{j l}\}}(\{\la\})=&
\pl_{l=1}^m\pl_{k=1}^{2s-1}\left(\pl_{j=1}^{\a_l-1}
(\la_{a_{j l}}-\xi_l+i(s-k))
\pl_{j=1}^{2s+1-\a'_l}
(\la_{a'_{j l}}-\xi_l+i(s-k+1))\right)
\times\nonumber\\
&\times \pl_{n\le l} G_{l n}^{DD}(\{a_{jl},a_{jn}\})
 G^{DA}_{l n}(\{a_{jl},a'_{jn}\})
 G^{AD}_{l n}(\{a'_{jl},a_{jn}\}) G^{AA}_{l n}(\{a'_{jl},a'_{jn}\})
%\times&\,\, 
\label{underintm}
\end{align}
Where the ``two sites'' contributions $G_{l n}$ for
 two different sites $l$ and $n$ are
\begin{align}
G_{l n}^{DD}=&
\frac{\pl_{j=1}^{\a_l-1}
\big(\la_{a_{jl}}-\xi_n-is\big)
\pl_{k=1}^{\a_n-1}
\big(\la_{a_{k n}}-\xi_l+is\big)}{\pl_{j=1}^{\a_l-1}
\pl_{k=1}^{\a_n-1}\big(\la_{a_{jl}}-\la_{a_{kn}}- i
\big)},\\
G_{l n}^{DA}=&\frac{\pl_{j=1}^{\a_l-1}
\big(\la_{a_{jl}}-\xi_n-is\big)
\pl_{k=1}^{2s-\a_n+1}
\big(\la_{a'_{k n}}-\xi_l-i(s-1)\big)}{\pl_{j=1}^{\a_l-1}
\pl_{k=1}^{2s-\a'_n+1}\big(\la_{a_{jl}}-\la_{a'_{kn}}- i
\big)},\\
%\begin{align}
G_{l n}^{AD}=&
\frac{\pl_{j=1}^{2s-\a'_l+1}
\big(\la_{a'_{jl}}-\xi_n+i(s+1)\big)
\pl_{k=1}^{\a_n-1}
\big(\la_{a_{k n}}-\xi_l+is\big)}{\pl_{j=1}^{2s-\a'_l+1}
\pl_{k=1}^{\a_n-1}\big(\la_{a_{kn}}-\la_{a'_{jl}}- i
\big)},\\
G_{l n}^{AA}=&\frac{\pl_{j=1}^{2s-\a'_l+1}
\big(\la_{a'_{jl}}-\xi_n+i(s+1)\big)
\pl_{k=1}^{2s-\a_n+1}
\big(\la_{a'_{k n}}-\xi_l-i(s-1)\big)}{\pl_{j=1}^{2s-\a'_l+1}
\pl_{k=1}^{2s-\a'_n+1}\big(\la_{a'_{kn}}-\la_{a'_{jl}}- i
\big)},
\end{align}
and the diagonal terms are given by
\begin{align}
G^{DD}_{ll}G^{AA}_{ll}=&\pl_{j,k=1\atop{j>k}}^{\a_l-1}
\frac{%(\la_{a_{j n}}-\xi_k+is\e_j^\b)
%(\la_k^\g-\xi_j+is\e_k^\g)
\la_{a_{j l}}-\la_{a_{k l}}}{(\la_{a_{j l}}-\la_{a_{k l}})^2+1}
\pl_{j,k=1\atop{j>k}}^{2s-\a'_l+1}
\frac{\la_{a'_{j l}}-\la_{a'_{k l}}}{(\la_{a'_{j l}}-\la_{a'_{k l}})^2+1},\\
G^{AD}_{ll}G^{DA}_{ll}=&\pl_{j=1}^{\a_l-1}\pl_{k=1}^{2s-\a'_l+1}
\frac 1{\la_{a_{jl}}-\la_{a'_{kl}}- i}
\end{align}
 
 We obtained here the algebraic part of the expression for a
 $m$-point correlation function
for an arbitrary spin Heisenberg chain.
Considering the determinant part we can easily see using
 the same arguments as in one-point  
case that
the main proposition about ``the substrings'' 
(the terms in the sum (\ref{mpointsum})
with  $a'_{j l}>N$ produce  non zero contributions  if
$a'_{j l}=N+2sk$ $k<l$ or if there is $a'_{k l}=a'_{j l} +1$ with $k<j$)
 is also valid in the $m$ point case. 
It means 
that the determinants are always simple and contain only  Gaudin columns and 
columns $p_a'(\la_b-\xi_k)$.  More precisely:
 every term of the sum (\ref{mpointsum}) contain a 
determinant which is obtained from the scalar product
\[\bra{0}\pl_{l=1}^M\pl_{j=1}^{k_l}
C(\xi_l-i(s-j+1))\!\!\!\pl_{b=k_1+\dots+k_m+1}^{N}\!\!\!
C(\la_b)\,\pl_{a=1}^N B(\la_a)\ket{0},\]
where $k_l=a'_{j_{\mathrm{min}} l}-1$, the ``substring end''
 $a'_{j_{\mathrm{min}} l}$ being the minimal $a'_{j l}>N$ in the
corresponding term. The corresponding matrix appearing in the sum (\ref{mpointsum})
has the following columns
\begin{alignat*}{2}
\tilde{\Psi}_{a b}=&\Phi'_{ab},\qquad &b>\sul_{l=1}^m k_l&,\\
\tilde{\Psi}_{a b}=&\frac 1{(\la_a-\xi_l+i(s-j+1))
(\la_a-\xi_l+i(s-j))}= p_j'(\la_a-\xi_l),\qquad &\sul_{r=1}^{l-1}
 k_r <b\le \sul_{r=1}^{l} k_r &,
\end{alignat*} 
where $j= b-\sul_{r=1}^{l-1} k_r$.

Thus we have a representation for the correlation functions 
for  a finite arbitrary spin Heisenberg chain. Being in some sense very similar
to their spin $\frac 12$ counterparts these representations are rather 
complicated for big spins. For this reason in the next sections we will  consider
only the first generalisation of the results of \cite{kmt2} which is the spin 1
chain. 

The next step  of our approach is  the thermodynamic limit for the ground state. 
On this stage the main 
difference with the spin $\frac 12$ case appears as the ground state is constructed
 of bound
 states ($2s$-strings).
It produces some new difficulties which will be considered in
 the next section using the simplest 
example of the one point functions.

\section{One point functions}
\setcounter{equation}{0}
To illustrate the last stage of the calculation, i.e. the
introduction of the string solution of the Bethe equations for the 
ground state we begin with the simplest example,
namely with the one point functions $f_k(1)$ (corresponding to
the diagonal elementary matrices $E^{kk}$)

Consider the simplest correlation function (1-point emptiness formation
 probability) 
of the spin 1 XXX chain
 in the homogeneous case:
\begin{equation}
\label{def1}
f_{3}(1)=\bra{vac}(\tau_2^{(2)})^{-1}(-i)D(0)D(-i)\ket{vac}
\end{equation}
Using the action of the operators $D$ on the vacuum and the scalar product formula 
we easily obtain (there is no difference with the case spin 1/2) for the
 finite chain:
\begin{equation}
f_{3}(1)=i \sum_{a}\sum_{b\neq a}
\frac{\la_a\la_b}{\la_a-\la_b-i}\,\,\frac{\det\Psi}
{\det\Phi'}
 \end{equation}
Where $\Phi'$ is the Gaudin matrix :
 \begin{equation}
\Phi'_{jk}=(M K(\la_j)-\sul_{l}K(\la_j-\la_l))\delta_{jk}+K(\la_j-\la_k)
\end{equation}
with
\[K(\la)=\frac 2{(\la+i)(\la-i)}\]
and the matrix $\Psi$ is obtained from the scalar product:
 \begin{align}
\Psi_{jk}=&\Phi'_{jk},\qquad j\neq a,b\nonumber\\
\Psi_{ak}=&\frac 1{\la_k(\la_k-i)}\equiv p'_+(\la_k)\nonumber\\
\Psi_{bk}=&\frac 1{\la_k(\la_k+i)}\equiv p'_-(\la_k)\nonumber.
\end{align}
Hence one can again divide one matrix by another 
and reduce it to a $2\times 2$ matrix:
 \begin{align}
({\Phi'}^{-1}\Psi)_{jk}=&\delta_{j k},\qquad j\neq a,b\nonumber\\
({\Phi'}^{-1}\Psi)_{ak}=& \phi_k^+\nonumber\\
({\Phi'}^{-1}\Psi)_{bk}=& \phi_k^-,\nonumber
\end{align}
where $\phi_{k}^{\pm}$ are solution of the following systems of 
linear equations:
\begin{equation}
\label{le1}
(M K(\la_k)-\sul_{j}K(\la_j-\la_k))\phi_{k}^\pm+\sul_{j}K(\la_j-\la_k))
\phi_{j}^\pm=p'_\pm(\la_k)
\end{equation}
Thus the 1-point function (\ref{def1}) is given by
\begin{equation}
\label{corr1}
f_{3}(1)= i\sum_{a}\sum_{b\neq a}\frac{\la_a\la_b}{\la_a-\la_b-i}\det\left(
\ba{cc}\phi_a^+&\phi_b^+\\ \phi_a^-&\phi_b^-\ea\right)
 \end{equation}

In the thermodynamic limit the ground state of the spin 1 
XXX chain is built of 2-strings
\[M\rightarrow\infty:\quad\la_{2k-1}\rightarrow\mu_k+\frac i2,
\quad\la_{2k}\rightarrow\mu_k-\frac i2, \quad \mathrm{Im}(\mu_k)=0\]
To obtain the  equations for the analytic part in the thermodynamic limit
 one should take into account the finite size corrections to this 
string picture as some terms in  (\ref{le1}) become singular. To 
 analyse the  excited states with finite energy of the XXX spin $\frac 12$ one usually considers first 
the string limit and only then the thermodynamic limit, as these corrections are exponentially
small. However for the ground state of the spin $1$ XXX model one cannot  use this 
method in a rigorous way\footnote[1]{this method leads to the same result for the correlation
functions as one described below but in a more complicated way} as the corrections to the string
picture calculated in   
  \cite{dVW,kb,kbp,suz} are of the
order $\frac 1M$: 
%Note that  one can obtain from the Bethe equation for fixed $\mu$ that
\[\la_{2k-1}-\la_{2k}-i=2i\frac{\a_k}M+o(\frac 1M),\qquad \a>0 \]
%also one should note that distance between the neighbour string centres is 
%\[\mu_k-\mu_{k-1}=O(\frac 1M).\]
  The correction $\a$ is always positive and it makes possible to rewrite the system 
 of  linear equations (\ref{le1})  as integral equations in the
thermodynamic limit  with a special choice of the integration contours 
near the singular point: 
\begin{align}
\varphi^\pm_1(\mu)+\!\!\int\limits_{-\infty}^\infty 
\! d\la\, K(\mu-\la)\varphi^\pm_1(\la)+\!\!\int\limits_{-\infty}^\infty 
\! d\la\, K(\mu-\la+i+i0)\varphi^\pm_2(\la)=&p'_\pm(\mu+\frac i2),\nonumber\\
\varphi^\pm_2(\mu)+\!\!\int\limits_{-\infty}^\infty 
\! d\la\, K(\mu-\la)\varphi^\pm_2(\la)+\!\!\int\limits_{-\infty}^\infty 
\! d\la\, K(\mu-\la-i-i0)\varphi^\pm_1(\la)=&p'_\pm(\mu-\frac i2)\label{ie},
\end{align}
 where $\varphi^\pm_1(\mu_k)\equiv M\rho(\mu_k)\phi^\pm_{2k-1}$ and 
      $\varphi^\pm_2(\mu_k)\equiv M\rho(\mu_k)\phi^\pm_{2k}$.  
 We used here the integral equation for density
of strings in the ground state 
\[\rho(\la)=\frac 1{2\cosh(\pi\la)}.\]
The solution of this system can be easily obtained
\begin{alignat}{2}
\varphi^+_1(\la)&=\rho(\la), \qquad&\varphi^+_2(\la)&=0,\nonumber\\
\varphi^-_1(\la)&=0, \qquad&\varphi^-_2(\la)&=\rho(\la).  \label{solie}
\end{alignat}

Now we can substitute this result to the expression for the one-point function (\ref{corr1}), 
replacing sums by integral avoiding the singular point in the {\it same way}  as in the integral equation
(taking into account the sign of the finite size corrections)
(\ref{ie}).
\begin{equation}
f_{3}(1)=\frac i4
\int\limits_{-\infty}^\infty d\la\int\limits_{-\infty}^\infty d\mu
\frac{1}{\cosh(\pi\la)\cosh(\pi\mu)}\left( \frac{(\la+\frac i2)(\mu-\frac i2)}{\la-\mu+
i0}- \frac{(\la-\frac i2)(\mu+\frac i2)}{\la-\mu-
2i}
\right)
\end{equation}
One of these two integrals can be calculated as  only the pole in the point $\mu=\la$
contributes. We obtain finally:
%Now we can calculate the only contribution to the 1-point function:
\begin{equation}
f_{3}(1)=
\frac \pi 2\int\limits_{-\infty}^\infty d\la\,\frac{\la^2+\frac 14}
{\cosh^2(\pi\la)}=
\frac 13.
\end{equation}
Of course this result can be obtained directly from the symmetry of the model, but
this calculation illustrate well how to deal with strings in our method 
and it can be useful not only for the  more general case of $m$ point functions
 which will be
considered in the next section but also for the 
computation of more general correlation functions,
depending, for example, on the temperature.
  
Two other one point functions can be calculated in a 
rather similar way but here we should consider also the
 ``$A$-type'' sums which contain more terms than 
``$D$-type'' sums considered in the previous example.
 We will show, using the simplest example of 
the one point functions $f_2(1)$ and $f_1(1)$ that this 
problem can be solved exactly as in the spin
$\frac 12$ case by moving the corresponding contour of integration.

Consider the function 
\[f_2(1)=\frac 12\bra{vac}\tau_2^{-1}(-i)\Big(A(0)D(-i)+
D(0)A(-i)+C(0)B(-i)+B(0)C(-i)\Big)\ket{vac}\]
We easily obtain a finite lattice representation for this function (\ref{onepsumm})
\begin{equation}
f_{2}(1)=-2i \sul_{a=1}^N\sul^N_{b=1\atop{b\neq a}}
\frac{\la_a(\la_b+i)}{\la_a-\la_b-i}\,\det\left(
\ba{cc}\phi_a^+&\phi_b^+\\ \phi_a^-&\phi_b^-\ea\right)
+2\sul_{a}\frac{\la_a}{\la_a-  i}\phi_a^-
\end{equation}
One should mention that the sum over index $a$ is a
 $D$-type sum and the sum over index $b$ is
a $A$-type sum and it contains one additional term 
(only one because of the substring limitation).

As in the previous case we can rewrite these sums 
as integrals in the thermodynamic limit: 
\begin{align}
f_{2}(1)=-\frac i2&\int\limits_{-\infty}^{\infty}d\la
\int\limits_{-\infty}^{\infty} d\mu
 \frac{1}{\cosh(\pi\la)\cosh(\pi\mu)}\left( \frac{(\la+\frac i2)(\mu+\frac i2)}{\la-\mu+
i0}- \frac{(\la-\frac i2)(\mu+\frac {3i}2)}{\la-\mu-
2i}
\right)+\nonumber\\
+&\int\limits_{-\infty}^{\infty} d\la \frac{1}{\cosh(\pi\la)}
\frac{\la-\frac i2}{\la_a-3\frac i2}
\label{4+2}
\end{align}
One should note that function $\rho(\la)=\frac{1}{2\cosh(\pi\la)}$ is the same as in the 
spin $\frac 12$ case and has a pole 
at
 $\la=-\frac i2$   and its residue is 
\[2\pi i \left.\mathrm{Res} \rho(\la)\right|_{\la=-\frac i2}=-1.\]
It means in particular that shifting the contour of
 integration on the variable $\mu$ to
the line parallel to the real axis with $\mathrm{Im}(\mu)=-1$
 one crosses the pole of $\rho(\la)$
for the second double integral in (\ref{4+2}) (for the first double integrals
 one can move the contour
without crossing any poles) and the contribution of the pole is
 exactly the  single 
integral in (\ref{4+2}). Finally we get
\begin{equation*}
f_{2}(1)=\frac i2\int\limits_{-\infty}^{\infty}d\la
\int\limits_{-\infty}^{\infty} d\mu 
\frac{1}{\cosh(\pi\la)\cosh(\pi\mu)}\left( \frac{(\la+\frac i2)(\mu-\frac i2)}{\la-\mu+
i}- \frac{(\la-\frac i2)(\mu+\frac {i}2)}{\la-\mu-
i}\right)
\end{equation*}
Here once again one can reduce this expression to a single integral: 
\begin{align}
f_{2}(1)=\frac 
\pi 2\int\limits_{-\infty}^\infty d\la\,\frac{\frac 12-2\la^2}
{\cosh^2(\pi\la)}=
\frac 13.
\end{align}

Considering the last one point function $f_1(1)$ one can obtain 
in a similar way that 
one should move both contours to obtain the same double integral representation as for  $f_3(1)$:
\begin{equation}
f_{1}(1)=\frac \pi 2\int\limits_{-\infty}^\infty 
d\la\,\frac{\la^2+\frac 14}{\cosh^2(\pi\la)}=
\frac 13.
\end{equation}

In this section we considered the simplest examples 
of the correlation functions, but however
this simple example illustrates quite well the basic 
properties of  the thermodynamic limits
for the ground state constructed of the 2-strings.
 We have also shown that the $A$-type sums should be replaced by the integrations
 over  shifted contours 
(as in spin $\frac 12$ case).

In the last section we show how this technique can 
be used for the general $m$-point
functions for the spin 1 Heisenberg XXX chain. The proofs of them 
in general are absolutely 
equivalent to the calculations in this section, but contain some 
very cumbersome formulae which
we omit in general for the intermediate steps. Also this method can be used for arbitrary
spin and we give in the end one of the possible multiple integral representations for 
the correlation functions of the higher spin chains.

\section{Correlation functions}
\setcounter{equation}{0}

In this section we generalise  the  results of the previous section
for the general $m$-point
equal-time correlation 
functions of the XXX chain spin 1 in the thermodynamic limit.

We calculate the following correlation functions or, more precisely, the 
elementary blocks which permit to construct any $m$ point correlation 
function:

\begin{equation}
\label{def}
f_{\{\a,\a'\}}(m)=\frac {\bra{\psi_g}\pl_{j=1}^m 
E^{\a'_j,\a_j}_j\ket{\psi_g}}{\bra{\psi_g}\psi_g\r}
\end{equation}
where $E^{\a'_j,\a_j}_j$ are elementary local $3\times 3$
matrices $E^{\a',\a}_{l k}=\delta_{l,\a'}\delta_{k,\a}$ and $|\psi_g\r$ 
is the ground state of 
the model (in the spin 1 case the number of quasiparticle $N$ in the ground
 state is equal to
the number of sites $M$).

For a finite spin 1 chain we obtained (\ref{mpointsum}) that this correlation function can be 
represented as multiple sums:
\begin{equation}
f_{\{\a,\a'\}}(m)=(-1)^{\sum(\a'_l-1)}
\frac{\left(\pl_{l=1}^m C_{2}^{\a_l-1}C_{2}^{\a'_l-1}\right)^{\frac 12}}{
 i^m \pl_{j,k=1\atop{j>k}}^m
(\xi_j-\xi_k)^2((\xi_j-\xi_k)^2+1)}\sul_{\{a_j,a'_j\}} \mathbf{H}_{\{a,a'\}}
(\{ \la \})
\det_{2m} \tilde{S}(\{a,a'\}).
\end{equation}
In the thermodynamic limit the solution of the Bethe equations 
corresponding to the ground state  consists of $2$-strings distributed with 
the following density: 
\[\rho_{\mathrm{tot}}(\la)=\frac 1M\sul_{n=1}^M\rho(\la-\xi_n),\]
As in the spin $\frac 12$ case  it is convenient to introduce a 
set of indices $b_k$  which is the set of $a_{j l}, a'_{j l}$ ordered in a special way: 
\[\{b_1,\dots, b_{2m}\}=\{\{a'_{j m}\},\dots, \{a'_{j 1}\},
\{a_{j 1}\},\dots,\{a_{j m}\}\},\]
with local subsets $\{a_{j 1}\}$, $1\le j\le\a_l-1$  and $\{a'_{j l}\}$,
 $1\le j\le 3-\a'_l$,
(these subsets   can  contain  one, two  or no elements).

Let us now consider the determinant part of the expression 
 (\ref{mpointsum}). In the 
thermodynamic limit we can divide the matrix in the numerator 
by the Gaudin matrix, or more 
precisely we can calculate $\det({\Phi'}^{-1}\Psi(\{a,a'\}))$. 
This determinant can
be written as a determinant of a   $2m\times 2m$ matrix
and 
the matrix elements of this matrix are given by the
 inhomogeneous version the integral equations (\ref{ie}) with $p_{\pm}(\mu-\xi_j\pm \frac i2)$
in the r.h.s. Due to the translation invariance of this equations we obtain just the same solution 
with a shift $\varphi_{1,2}^\pm(\mu-\xi_j)$. Finally we obtain: 
\begin{alignat}{2}
\det_M({\Phi'}^{-1}\Psi(\{a,a'\}))=&\det_{2m}\tilde{S}(\{a,a'\}),&\nonumber \\
\tilde{S}_{j k}(\{a,a'\})=& -\delta_{b_{j}-M,k}, \quad &b_j>M,\nonumber\\
\tilde{S}_{j 2k-1}(\{a,a'\})=& \frac{1+(-1)^{b_j}}2 
\frac{\rho(\la_{b_j}-\xi_k-\frac i2)}
{\rho_{\mathrm{tot}}(\la_{b_j}-\frac i2)},\quad &b_j\le M,\nonumber\\
\tilde{S}_{j 2k}(\{a,a'\})=& \frac{1-(-1)^{b_j}}2 
\frac{\rho(\la_{b_j}-\xi_k+\frac i2)}
{\rho_{\mathrm{tot}}(\la_{b_j}+\frac i2)},\quad &b_j\le M.
\label{anpart}
\end{alignat}

Now all the sums over $b_l$ from $1$ to $M$ can be replaced by
 integrals taking into account
account that we obtain a sum of two integrals obtained from $b_l=2j,\quad j=1,\dots M/2$ and
 $b_l=2j-1,\quad j=1,\dots M/2$ dealing with the contours near the singularities of the algebraic
part in the same way as in the previous section. 
 Replacing sums  by integrals we use the following rules and notations:
\begin{alignat*}{2}
\sul_{b_l=2j-1\atop{j=1}}^{M/2}
 \longrightarrow&\int\limits_{-\infty+\frac i2+i0}^{\infty+\frac i2+i0}
d \la_l\, \rho_{\mathrm{tot}}(\la_l-\frac i2)
\qquad&\sul_{b_l=2j\atop{j=1}}^{M/2}
 \longrightarrow&\int\limits_{-\infty-\frac i2-i0}^{\infty-\frac i2-i0}
d \la_l\, \rho_{\mathrm{tot}}(\la_l+\frac i2)\\
\sul_{a_{ k r}=2j-1\atop{j=1}}^{M/2}
 \longrightarrow&\int\limits_{-\infty+\frac i2+i0}^{\infty+\frac i2+i0}d \nu_{ k r} \,
\rho_{\mathrm{tot}}(\nu_{k r}-\frac i2)
\qquad&\sul_{a_{k r}=2j\atop{j=1}}^{M/2}
 \longrightarrow&\int\limits_{-\infty-\frac i2-i0}^{\infty-\frac i2-i0}d \nu_{k r} \,
\rho_{\mathrm{tot}}(\nu_{k r}+\frac i2)\\
\sul_{a'_{k r}=2j-1\atop{j=1}}^{M/2}
 \longrightarrow&\int\limits_{-\infty+\frac i2+i0}^{\infty+\frac i2+i0}d \nu'_{k r}\, 
\rho_{\mathrm{tot}}(\nu'_{k r}-\frac i2)
\qquad
&\sul_{a'_{k r}=2j\atop{j=1}}^{M/2} \longrightarrow
&\int\limits_{-\infty-\frac i2-i0}^{\infty-\frac i2-i0}d \nu'_{k r}\, 
\rho_{\mathrm{tot}}(\nu'_{k r}+\frac i2).
\end{alignat*}
In this notation the set $\{\la_l\}$ is the same set as $\{\nu_{k r},
 \nu'_{k r}\}$ but
ordered in a special way: 
\[\{\la_1,\dots, \la_{2m}\}=\{\{\nu'_{k m}\},\dots, \{\nu'_{k 1}\},
\{\nu_{k 1}\},\dots,\{\nu_{k m}\}\}.\]
 We will use also the following parameters:
\begin{alignat*}{2}
\varepsilon_l=&\frac 12,\qquad &\text{if}  \quad &\la_l=\nu_{k r} 
\quad (D-\mathrm{type}),\\
\varepsilon_l=&-\frac 12,\qquad &\text{if}  \quad &\la_l=\nu'_{k r}
\quad (A-\mathrm{type}),
\end{alignat*}
we can also associate to every $\la_l$ the corresponding site number $r_l$ if
$\la_l=\nu_{k r_l}$ or $\la_l=\nu'_{k r_l}$. This notation are very useful
to simplify our formulae.

Replacing the sums by two integrals one should mention that for different contours for $\la_j$
we obtain different functions in the determinant, namely $\varphi_1^\pm(\la_j-\xi_k-\frac i2)$
for the upper contour and  $\varphi_2^\pm(\la_j-\xi_k+\frac i2)$ for the lower one. To simplify the
formulae  we introduce the following function: 
\begin{alignat}{2}
\phi^\pm(\la)=&\varphi_1^\pm(\la-\frac i2),\quad &\mathrm{Im}(\la)>&-\frac 12,\nonumber\\
\phi^\pm(\la)=&\varphi_2^\pm(\la+\frac i2),\quad &\mathrm{Im}(\la)<&-\frac 12.
\end{alignat}
(Note that we can deal with this function as with an analytic function if the integration contours
do not cross the line $\mathrm{Im}(\la)=-\frac 12$).

We should now analyse the terms with $a'_{k r}>M$. As in the spin $\frac 12$ case they 
can be written as integrals around the corresponding poles  of the determinant
part in the points $\la_j=\xi_k$ (for the upper contour), $\la_j=\xi_k-i$ (for the lower contour).
 More precisely we can replace the complete sum over $a'_{k r}>M$ by the following
integral:
\begin{align*}\sul_{a'_{k r}=2j} &\longrightarrow
\left(\int\limits_{-\infty-\frac i2-i0}^{\infty-\frac i2-i0}+\sul_{l=1}^m\oint_{\Gamma'_l}\right)
d \nu'_{k r}\, 
\rho_{\mathrm{tot}}(\nu'_{k r}+\frac i2)\\
\sul_{a'_{k r}={2j-1}} &\longrightarrow
\left(\int\limits_{-\infty+\frac i2+i0}^{\infty+\frac i2+i0}+\sul_{l=1}^m\oint_{\Gamma_l}\right)
d \nu'_{k r}\, 
\rho_{\mathrm{tot}}(\nu'_{k r}-\frac i2).\end{align*}
where $\Gamma_l$ and $\Gamma'_l$ are small contours around the points $\xi_l$ and $\xi_l-i$. 
It means that as in the spin $\frac 12$ case the contribution of the poles can be absorbed 
into the integrals by moving all the contours for $\nu'_{k r}$  ($A$-type variables) down by $i$
(to avoid crossing of some additional poles one should move first all the lower contours and than 
the upper ones).

Now we can write  a multiple integral representation for the correlation functions
of the spin 1 XXX chain:
\begin{align}
&f_{\{\a,\a'\}}(m)=%(-1)^{\sum(\a'_l-1)}
i^m 
\frac{\left(\pl_{l=1}^m C_{2}^{\a_l-1}C_{2}^{\a'_l-1}\right)^{\frac 12}}{
 \pl_{j,k=1\atop{j>k}}^m
(\xi_j-\xi_k)^2((\xi_j-\xi_k)^2+1)}\times \nonumber\\
&\times\left(\int\limits_{-\infty-\frac i2-i0}^{\infty-\frac i2-i0}
+\int\limits_{-\infty+\frac i2+i0}^{\infty+\frac i2+i0}\right)
d\la_1\dots  \!\!\!\left(\int\limits_{-\infty-\frac i2-i0}^{\infty-\frac i2-i0}
+\int\limits_{-\infty+\frac i2+i0}^{\infty+\frac i2+i0}\right) d\la_{2m}\,\,
\mathbf{H}_{\{\a,\a'\}}(\{\la_l\})\det_{2m} \mathcal{S}(\{\la\}).
\label{result2a}
\end{align}
The  algebraic part for spin 1 can be written as:
\begin{align}
\mathbf{H}_{\{\a,\a'\}}(\{\la_l\})=\pl_{l=1}^{2m}
\pl_{k=1}^m(\la_l-\xi_k)
\frac{\pl_{l=1}^{2m}
\left(\pl_{k=1}^{r_l-1}(\la_l-\xi_k-2i\varepsilon_l)\pl_{k=r_l+1}^{m}
(\la_l-\xi_k+2i\varepsilon_l)
\right)}{\pl_{l>n}\left(\vphantom{\pl_{k=1}^{r_l-1}}
\la_l-\la_n-i(\varepsilon_l+\varepsilon_n)^2\right)},
\label{apart}
\end{align}
and the $2m\times 2m$ matrix $\mathcal{S}$ is defined as
\begin{align}
\mathcal{S}_{j,2k-1}=&\phi^-(\la_j-\xi_k),\nonumber\\
\mathcal{S}_{j, 2k}=&\phi^+(\la_j-\xi_k).
\end{align}

This is one of many possible forms of the results which can be easily generalised to the 
higher spin cases. One should note that the correlation functions are written once again as a 
multiple integral (or a sum of multiple integrals) and the integrals are taken over
the solution of Bethe equations for the ground state ($2$-strings here). The expression
under the integral can be once again  separated into two distinctive parts: one defined
by the choice of local operators (algebraic part) and one defined by the ground state (determinant).

This result can be also rewritten in many different forms. First of all as  for the one-point
functions one can reduce the number of integration and obtain the result only as $m$ integrals
(this is a particularity of the spin $1$ case, it cannot be done for the higher spins). First
of all one should note that $S_{j,2k-1}=0$ for upper integration contours and $S_{j,2k}=0$ for
lower ones (\ref{solie}). It means that the determinant has a block diagonal structure and
the integrals can be rewritten as a sum of integrals 
with $m$ integrals over the upper contours  and $m$ integrals over lower ones:
\begin{align}
&f_{\{\a,\a'\}}(m)=
\frac{%(-1)^{\sum(\a'_l-1)}
i^m \left(\pl_{l=1}^m C_{2}^{\a_l-1}C_{2}^{\a'_l-1}\right)^{\frac 12}}{
 \pl_{j,k=1\atop{j>k}}^m
(\xi_j-\xi_k)^2((\xi_j-\xi_k)^2+1)} \sul_{\{\la\}=\{\mu\}\cup\{\mu'\}}
\!\!(-1)^{[\s]}\!\!\int\limits_{-\infty-\frac i2-i0}^{\infty-\frac i2-i0}\!\!\!\!d\mu_1\dots
\int\limits_{-\infty-\frac i2-i0}^{\infty-\frac i2-i0}\!\!\!\!d\mu_m\times\nonumber\\
&\times
\int\limits_{-\infty+\frac i2+i0}^{\infty+\frac i2+i0}\!\!\!\!d\mu'_1\dots
\int\limits_{-\infty+\frac i2+i0}^{\infty+\frac i2+i0}\!\!\!\!d\mu'_m\,\,
\mathbf{H}_{\{\a,\a'\}}(\{\la_l\})\det_{m} \mathcal{W}(\{\mu+\frac i2\})\mathcal{W}(\{\mu'-\frac i2\}).
\label{result2b}
\end{align} 
where the sum is taken over all possible partitions of the set $\{\la\}$
with $2m$ elements into two subsets with $m$ elements and $[\s]$ is the sign of the following
 permutation
\[ \s({\{\la_1,\dots,\la_{2m}\})=\{\mu_1,\dots,\mu_m,\mu'_1,\dots,\mu'_m\}},\]
and $\det\mathcal{W}(\{\mu'\})$ is the spin $\frac 12$ determinant:
  \[\mathcal{W}_{j k}=\rho(\mu_j-\xi_k).\]
this representation is also convenient as it contains only $m\times m$ matrices and
only meromorphic functions. Now we can move the contours in this integral to obtain that 
only poles in $\la_j-\la_k+i(\varepsilon_j+\varepsilon_K)$ contribute, which gives 
a representation of the result as  a sum over all possible splitting of the set of $2m$ variables
 $\la_l$ into
$m$ pairs  which form strings
 $\la_l=\nu_j-\frac i2$,  $\la_l'=\nu_j+\frac i2$
 but every term contains only  $m$ integrals over the string centres $\nu_j$:
\begin{align}
f_{\{\a,\a'\}}(m)=&%(-1)^{\sum(\a'_l)}
(2\pi)^m 
\frac{\left(\pl_{l=1}^m C_{2}^{\a_l-1}C_{2}^{\a'_l-1}\right)^{\frac 12}}{
 \pl_{j,k=1\atop{j>k}}^m
(\xi_j-\xi_k)^2((\xi_j-\xi_k)^2+1)}\int\limits_{-\infty}^{\infty}d\nu_1\dots
\int\limits_{-\infty}^{\infty}d\nu_m
\,\,\det_m^2 \mathcal{W}(\{\nu\})\times\nonumber\\
&\times\sul_{\{1,2,\dots,2m\}=\cup_{j=1}^m\{l_j l'_j\}}G_{\{\a,\a'\}}(\{\nu\},
\{\{l_j l'_j\},
j=1\dots m\}).
\label{result1}
\end{align}

The algebraic part here is however much more complicated than in the spin $\frac 12$ 
case.
 It is written as a sum
of $(2m-1)!!\equiv 1\ \cdot\ 3\ \cdot\dots\ \cdot\ (2m-1)$ terms, where every 
term is
 the corresponding residue of the general algebraic part $\mathbf{H}_{\{\a,\a'\}}$
\begin{align}
&G_{\{\a,\a'\}}(\{\nu\},\{\{l_j l'_j\},j=1\dots m\})=\nonumber\\
&=\left.\mathrm{Res}\right|_{\la_{l_1}=\la_{l'_1}+i
(\varepsilon_{l_1}+\varepsilon_{l'_1})^2}\dots
\left.\mathrm{Res}\right|_{\la_{l_m}=\la_{l'_m}+i
(\varepsilon_{l_m}+\varepsilon_{l'_m})^2} 
\mathbf{H}_{\{\a,\a'\}}(\{\la_l\}).
\end{align}

This is another possible form for the final result. It is rather particular as
the correlation functions are represented as $m$-integrals as in the spin $\frac 12$
case and the analytic part is just the square of the analytic part for spin 
 $\frac 12$.
It is also important to note that the homogeneous limit of this expression 
can be obtained 
exactly as in the spin $\frac 12$ case. 
\begin{equation}
\lim_{\xi_j\rightarrow 0
}\frac {\det^2_m  W(\{\mu\})}{
 \pl_{j,k=1\atop{j>k}}^m
(\xi_j-\xi_k)^2}= \det^2_m W^{\mathrm{hom}}(\{\mu\}), \qquad W^{\mathrm{hom}}_{j k}=
\frac 1{(k-1)!}\frac{\partial^{k-1}}{\partial\mu_j^{k-1}}\rho(\mu_j)
\label{homogen}
\end{equation}
Once again the analytic part will be just the
square of the corresponding spin $\frac 12$ algebraic part.

Thus we obtained several equivalent expressions for the correlation 
functions of the spin 1 
Heisenberg XXX chain. It is interesting to note that this result 
looks quite different in
comparison to the corresponding results obtained in \cite{Idz,BogW} 
(it is quite clear that for
the same quantity one can write many different integral 
representations and sometimes it
is rather difficult to prove that they are equivalent). 
The first (\ref{result2b}) result has a form which is much simpler, 
even if now we have $2m$ integrals
instead of $m$. Moreover similar results can be obtained for higher spins (\ref{mpointsum})
in the thermodynamic limit,
with more
complicated contours (in some sense we integrate always {\it over the strings}),
 determinant of a $2ms\times 2ms$ matrix  and corresponding general
algebraic part: 
\begin{align}
f_{\{\a,\a'\}}(m,s)
= &%(-1)^{\sum(\a'_l-1)}
\frac{\left(\pl_{l=1}^m C_{2s}^{\a_l-1}C_{2s}^{\a'_l-1}\right)^{\frac 12}}{
\left(\pl_{k=1}^{2s-1} i^k\, k!\right)^m \pl_{j,k=1\atop{j>k}}^m\pl_{r=1}^{2s}
\pl_{n=1}^{2r-1}
\Big(\xi_j-\xi_k-i(r-n)\Big)}\times\nonumber\\
&\times
\int\limits_{2s\text{-strings}}d\la_1 \dots \int\limits_{2s\text{-strings}}d\la_{2sm}\,\, 
\mathbf{H}^{(s)}_{\{\a,\a'\}}(\la_1,\dots,\la_{2sm})
\det_{2sm} \mathcal{S}^{(s)}(\{\la\}).
\label{mpointsint}
\end{align}
Here  the intergals are taken over the strings with corresponding little shifts:
\[\int\limits_{2s\text{-strings}}d\la\,\,
 f(\la)=\sum_{k=1}^{2s}\int\limits_{-\infty-(i+i0)(2s-k-\frac 12)}
^{\infty-(i+i0)(2s-k-\frac 12)}d\la \,\,\,f(\la),\]
the algebraic part is 
\begin{align}
\mathbf{H}^{(s)}_{\{\a,\a'\}}(\{\la_l\})=&\left(\pl_{p=1}^{2s-1}\pl_{l=1}^{2sm}
\pl_{k=1}^m(\la_l-\xi_k-i(s-p))\right)\times\nonumber\\
&\times\frac{\pl_{l=1}^{2sm}
\left(\pl_{k=1}^{r_l-1}(\la_l-\xi_k-2is\varepsilon_l)\pl_{k=r_l+1}^{m}
(\la_l-\xi_k+2is\varepsilon_l)
\right)}{\pl_{l>n}\left(\vphantom{\pl_{k=1}^{r_l-1}}
\la_l-\la_n-i(\varepsilon_l+\varepsilon_n)^2\right)},
\label{aparts}
\end{align}
and the $2sm\times 2sm$ matrix $\mathcal{S}^{(s)}$ is defined as
\begin{alignat}{2}
\mathcal{S}_{j,2s(k-1)+l}=&\rho(\la_j-\xi_k+2si-il+\frac i2),
 &\quad &-2s+l-1<\mathrm{Im}(\la_j)<-2s+l\nonumber\\
                     =& 0, &\quad &\text{otherwise}.
\end{alignat}
This representation can be proved in a very similar way to the spin $1$ case (\ref{result2b}).

We discussed in this paper the correlation functions of the 
higher spin XXX chains and we have
shown that the even if the ground state contains  bound 
states the correlation functions can be
calculated. We hope to use these result to calculate the 
mean values of local operators 
with respect
to any excited state for the spin $\frac 12$ Heisenberg 
chains and, hence, to obtain a 
representation for the
finite temperature correlation functions.

\section*{Acknowledgements}

I would like to thank J.-M. Maillet, V. Terras, N. Slavnov, A. Doikou
 and R. Weston for many interesting discussions
and comments. I'm also grateful to N. MacKay, E. Corrigan, P. Basillac and others
members of  the department of Mathematics in York 
for the interest in my work.
I would like also to thank one of the referees for
pointing out the results of the papers \cite{kbp,suz}.
This work was supported by UK EPSRC grant GR/M 73231.

\bibliographystyle{h-elsevier} %style de Nuclear Physics

\bibliography{biblio}

\begin{thebibliography}{10}

\bibitem{KitMT99}
N. Kitanine, J.M. Maillet and V. Terras,
\newblock Nucl. Phys. B 554 [FS] (1999) 647, math-ph/9807020.

\bibitem{IzeKMT99}
A.G. Izergin et~al.,
\newblock Nuclear Physics B 554 [FS] (1999) 679.

\bibitem{kmt2}
N. Kitanine, J.M. Maillet and V. Terras,
\newblock Nucl. Phys. B 567 [FS] (2000) 554, math-ph/9907019.

\bibitem{FadST79}
L.D. Faddeev, E.K. Sklyanin and L.A. Takhtajan,
\newblock Theor. Math. Phys. 40 (1979) 688.

\bibitem{mt}
J.M. Maillet and V. Terras,
\newblock Nucl. Phys. B 575 (2000) 627, math-ph/9911030.

\bibitem{JimMMN92}
M. Jimbo et~al.,
\newblock Phys. Lett. A168 (1992) 256.

\bibitem{JimM96}
M. Jimbo and T. Miwa,
\newblock Journ. Phys. A 29 (1996) 2923.

\bibitem{KulRS}
P.P. Kulish, N. Reshetikhin and E. Sklyanin,
\newblock Lett. Math. Phys. 5 (1981) 393.

\bibitem{takh}
L.A. Takhtajan,
\newblock Phys. Lett. 87A, (1982) 479.

\bibitem{bab}
H. Babujian,
\newblock Nucl. Phys. B 215 [FS7] (1983) 317.

\bibitem{FatZ}
A. Zamolodchikov and V. Fateev,
\newblock Yad. Fiz. 32 (1980) 581.

\bibitem{KirR}
A.N. Kirillov and N.Y. Reshetikhin,
\newblock Zap. Nauch. Semin. {LOMI} 145 (1985) 109.

\bibitem{Idz}
M. Idzumi,
\newblock (1993), hep-th/9307129.

\bibitem{BogW}
A.H. Bougourzi and R. Weston,
\newblock Nucl. Phys. B 417 (1994) 439.

\bibitem{Kon}
H. Konno,
\newblock Nucl. Phys. B 432 (1994) 457.

\bibitem{Gau67}
M. Gaudin,
\newblock La fonction d'onde de Bethe, Paris Masson (1983).

\bibitem{Kor82}
V.E. Korepin,
\newblock Commun. Math. Phys. 86 (1982) 391.

\bibitem{Sla89}
N.A. Slavnov,
\newblock Theor. Math. Phys. 79 (1989) 502.

\bibitem{dVW}
H.J. de Vega and F. Woynarovich, 
\newblock J. Phys A 23 (1990) 1613,

\bibitem{kb}
A. Kl\"{u}mper and M.T. Batchelor, 
\newblock J. Phys A 23 (1990) L189,

\bibitem{kbp}
A. Kl\"{u}mper, M.T. Batchelor and P.A. Pearce,
\newblock J. Phys A 24 (1991) 2341,

\bibitem{suz}
J. Suzuki,
\newblock J. Phys A 32 (1999) 2341,

\end{thebibliography}

\end{document}